\documentclass[pra,showpacs,preprintnumbers,amsmath,amssymb,tightenlines,twocolumn]{revtex4-1}

\usepackage{graphicx}
\usepackage{dcolumn}
\usepackage{bm}
\usepackage{epsfig}
\usepackage{amsmath}
\usepackage[export]{adjustbox}

\newcommand{\beginsupplement}{%
	\setcounter{table}{0}
	\renewcommand{\thetable}{A\arabic{table}}%
	\setcounter{figure}{0}
	\renewcommand{\thefigure}{A\arabic{figure}}%
	\setcounter{equation}{0}
	\def\theequation{A\arabic{equation}}
}

\newcommand{\ssection}[1]{{\noi  \it #1:}}
\newcommand{\bra}[1]{\langle\,{#1}\, |}
\newcommand{\ket}[1]{|\,{#1}\,\rangle}
\newcommand{\braket}[2]{\mbox{$\langle\,{#1}\, | \,{#2}\,\rangle$}}

\newcommand{\m}{\vspace{0.5cm}}

\newcommand{\sub}[2]{{#1}_{\mbox{\!\! \scriptsize #2}}}

\def\noi{\noindent}
\def\beq{\begin{equation}}
\def\eeq{\end{equation}}

\def\figurewidth{0.99}
\newcommand{\rref}[1]{ref.~\cite{#1}}
\newcommand{\fref}[1]{Fig.~\ref{#1}}
\newcommand{\frefp}[2]{Fig.~\ref{#1}~(#2)}

\newcommand{\eref}[1]{Eq.~(\ref{#1})}
\newcommand{\esref}[2]{Eqs.~(\ref{#1}) and (\ref{#2})}

\newcommand{\cref}[1]{chapter~\ref{#1}}

\newcommand{\Cref}[1]{Chapter~\ref{#1}}

\newcommand{\bref}[1]{(\ref{#1})}

\usepackage{ulem}  
\usepackage{xcolor}

\begin{document}
\title{Trapping and binding by dephasing}
\author{K.~Mukherjee}
\affiliation{Department of Physics, Indian Institute of Science Education and Research, Bhopal, Madhya Pradesh 462 066, India}
\author{S.~Poddar}
\affiliation{Department of Physics, Indian Institute of Science Education and Research, Bhopal, Madhya Pradesh 462 066, India}
\author{S.~W\"uster}
\affiliation{Department of Physics, Indian Institute of Science Education and Research, Bhopal, Madhya Pradesh 462 066, India}
\email{sebastian@iiserb.ac.in}
\begin{abstract}
Binding and trapping of particles usually rely on conservative forces, described by unitary quantum dynamics. We show that both can also arise solely from spatially dependent dephasing, the simplest type of decoherence. This can be based on continuous weak position measurements in only selected regions of space, for which we propose a practical realisation. For a single particle, we demonstrate a quantum particle-in-the-box based on dephasing. For two particles, we demonstrate their binding despite repulsive interactions, if their molecular states are dephased at large separations only. Both mechanisms are experimentally accessible, as we show for an example with Rydberg atoms in a cold gas background.
\end{abstract}

\maketitle

\ssection{Introduction}
%
Bound states due to conservative forces are central to our existence, combining fundamental particles into atoms, those into molecules and
the latter into complex biological structures. Many binding mechanisms arise from emergent rather than fundamental forces, and can 
be very weak, even while essential, such as when Cooper pairs bound by lattice-phonon mediated interactions cause superconductivity.
Quantum coherence is often a crucial element of the bound-state, as in the molecular states of the $H_2^+$ molecule, which change character from covalent binding to repelling depending on the relative phase of atomic orbitals \cite{book:brandsen}. In  larger systems, decoherence can become relevant \cite{book:schlosshauer,Schlosshauer_decoherence_review}, and while usually detrimental for quantum technologies \cite{Suter_RevModPhys}, it can also be a resource \cite{Poyatos_dissip_engineering_PhysRevLett,Verstraete_dissip_engineering,Vuglar_nonconsforces_PhysRevLett}.

Trapping and binding are two key requirements for complexity, and we show here that both can also arise due to the simplest type of decoherence: dephasing. Trapping in our proposal  exploits the quantum Zeno effect \cite{misra:quantumzeno,Kofman:antizeno}, as the position of a particle is inferred only \emph{outside} the trapping region \cite{Zhu:loss_zeno,Yan:loss_zeno_expt,Makrory:qmeasurement,Gordon:qlocalization}, and similarly binding relies on relative distance measurements that are only sensitive to large distances.
\begin{figure}[htb]
	\centering
	\epsfig{file=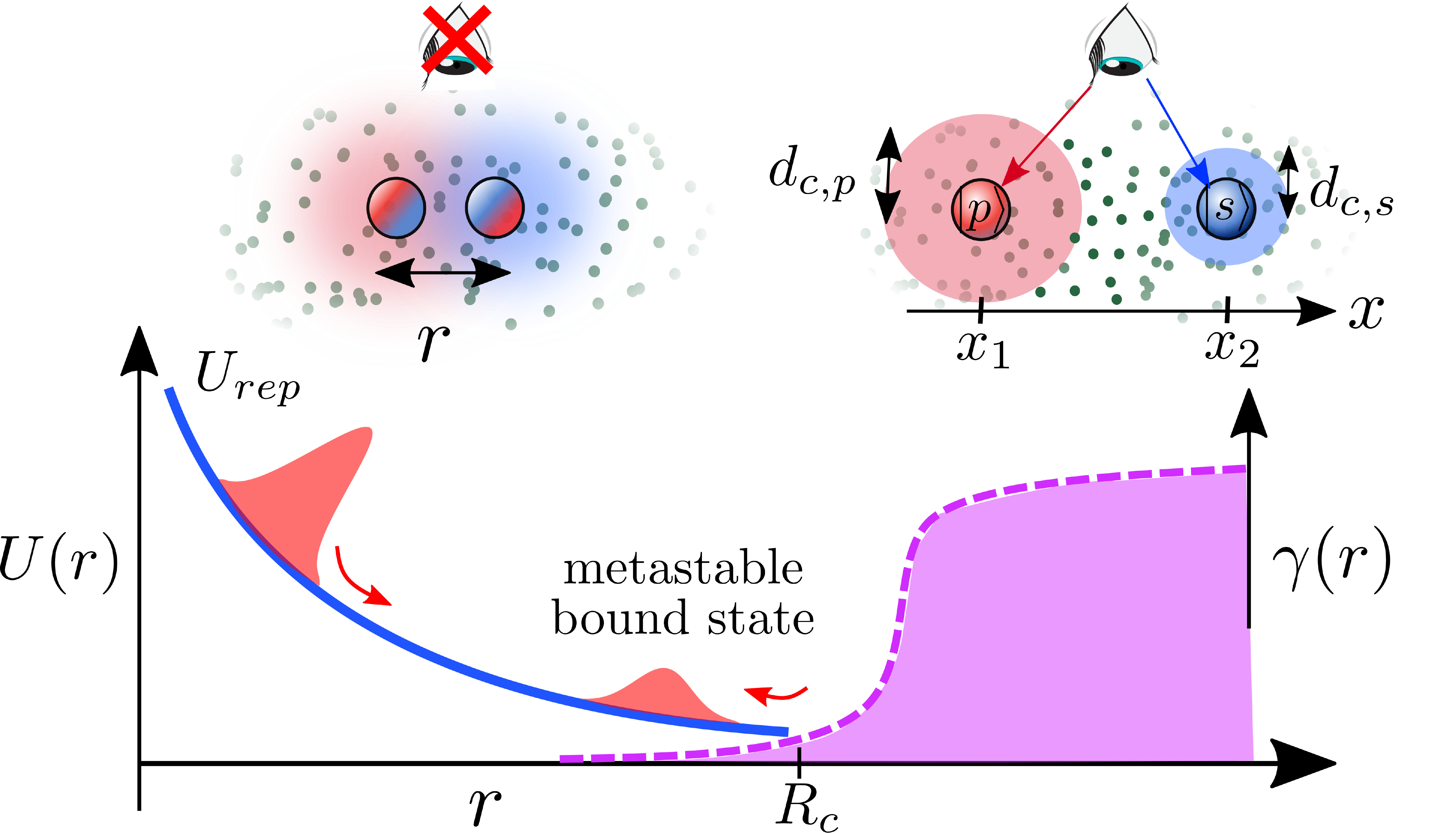,width= \linewidth}
	\caption{Selective relative distance measurements. A dimer of Rydberg atoms in $\ket{s}$ and $\ket{p}$ states (blue and red spheres) is separated by a distance $r$ and immersed in an ambient medium of ground-state atoms (green). (top) Electromagnetically induced transparency (EIT) in the medium can be used to measure position $x$ and state of dimer atoms at large separation (top right) $r=|x_2-x_1|$, whereas this breaks down at small $r$ (top left). (bottom) The nuclear wavefunction of the dimer (red) on a repulsive potential $\sub{U}{rep}(r)$ (blue) can then reflect off the steep dephasing profile $\gamma(r)$ (violet dashed) at large separation $(r\sim R_c)$, forming a metastable bound state.  \label{sketch_setup}}
\end{figure}
This enables a further class of decoherence processes, namely pure dephasing, as a tool for incoherent quantum state engineering \cite{Verstraete_dissip_engineering} of trapped states or bound states for quantum technologies. The emergence of such states was earlier shown only due to decoherence arising from loss processes and therefore involving dissipation as well \cite{Lemeshko:binding,Syassen:dissipation,Syassen:dissipation_exp,Ates:diss_binding,Cui:distance_dissi,Zhu:loss_zeno,Daley:3body_loss,Vuglar_nonconsforces_PhysRevLett}, which is not important here.

To begin, we summarize how the spatially selective position and internal state measurement of particles can affect their motion, described by an effective master equation. To be specific
we discuss the examples of a single Rydberg atom and a Rydberg dimer, for which a surrounding ultracold gas can provide the required position and state quantum non-demolition (QND) measurement, see \fref{sketch_setup}, but the principles can be ported to other platforms. Nextly, we construct an equivalent of the well known particle-in-the-box, but based on dephasing. In the final two sections we reach our main result, generalizing single particle trapping to a two-particle bound-state.

\ssection{Spatially selective dephasing}
\label{theory}
%
We first outline a general theory of $N$ Rydberg atoms embedded in an ultracold gas, which we later apply to a single Rydberg atom ($N=1$) or a Rydberg dimer ($N=2$),  as sketched in \fref{sketch_setup}. We let each Rydberg atom to be in states $\ket{s}=\ket{\nu s}$ and $\ket{p}=\ket{\nu p}$ with high principal quantum number $\nu$. The system can thus be in electronic states
$\ket{\pi_n} = \ket{ss\cdots p\cdots s}$, where the $n$'th atom is in $\ket{p}$ and the rest are in $\ket{s}$ \cite{wuester:review}. The effective Hilbert-space for the surrounding cold gas atoms is $\{\ket{g},\ket{e},\ket{u}\}$, predominantly in the electronic ground-state $\ket{g}$, but optically coupled via a low lying excited state $\ket{e}$ to an auxiliary Rydberg excited state $\ket{u}=\ket{\nu's}$, $\nu'\neq\nu$, in a ladder configuration resulting in Rydberg electromagnetically induced transparency (EIT) \cite{fleischhauer:review,friedler:longrangepulses,Mohapatra:coherent_ryd_det,mauger:strontspec,Mohapatra:giantelectroopt,schempp:cpt,sevincli:quantuminterf,parigi:interactionnonlin}. In this configuration the cold gas atoms only occupy the $\ket{u}$ state with a very low probability, hence we do not refer to them as Rydberg atoms, but as background atoms. We group all positions $\mathbf{X}_n$ of Rydberg atoms with latin indices $n$ into a collective coordinate $\mathbf{X}=[\mathbf{X}_1, \mathbf{X}_2, \cdots \mathbf{X}_N]^T$
and similarly for background atoms $\mathbf{x}=[ \mathbf{x}_1, \cdots  \mathbf{x}_{\sub{N}{bg}}]^T$ at position $\mathbf{x}_\alpha$, using greek indices for those.

The starting point for modelling is the many-body Master equation
 \begin{align}
 	\dot{\hat{\rho}}&=-i[\hat{H},\hat{\rho}]+\sum_\alpha \mathcal{L}_{\hat{L}_\alpha}[\hat{\rho}], 
 	\label{full_main}
 \end{align}
 where $\hat{H}$ is the total Hamiltonian, $\hat{\rho}$ the density matrix and $\mathcal{L}_{\hat{L}_\alpha}$ a superoperator $\mathcal{L}_{\hat{O}}=\hat{O}\hat{\rho}\hat{O}^\dagger-(\hat{O}^\dagger\hat{O}\hat{\rho}+\hat{\rho}\hat{O}^\dagger\hat{O})/2$, for spontaneous decay from $\ket{e}$ to $\ket{g}$ via the decay operator $\hat{O}=\hat{L}_\alpha=\sqrt{\Gamma_p} \hat{\sigma}_{ge}^{(\alpha)}$ with decay rate $\Gamma_p$. Here $\hat{\sigma}_{bb'}^{(\alpha)}=[\ket{b}\bra{b'}]_\alpha$ is acting on the $\alpha$'th background atom only. In \bref{full_main} we neglect motion for the moment, 
 thus $\hat{\rho}=\sum_{nm,\mathbf{b},\mathbf{b}'}\rho_{nm,\mathbf{b},\mathbf{b}'} \ket{\pi_n; \mathbf{b}}\bra{\pi_m; \mathbf{b}'}$, where
 $n$ and $m$ label the electronic states of the Rydberg atoms and vector $\mathbf{b}$ labels the background atoms with each entry in $b_\alpha \in \{g,e,u\}$. Since \bref{full_main} is intractable for many atoms, we now discuss three steps towards a simplified effective description.

{\it Step 1:} We assume the background gas settles into a steady state, enabling interaction enhanced imaging of states and positions of Rydberg atoms \cite{Gunter:EIT,guenter:EITexpt}. Its Hamiltonian for optical transitions is
\begin{eqnarray}
	\sub{\hat{H}}{EIT} &=& \frac{\Omega_p}{2} \hat{\sigma}_{eg}^{(\alpha)} + \frac{\Omega_c}{2} \hat{\sigma}_{ue}^{(\alpha)} + H.c. \nonumber \\
	&&- \Delta_p \hat{\sigma}_{ee}^{(\alpha)} - (\Delta_p + \Delta_c)\hat{\sigma}_{uu}^{(\alpha)}, \label{H_EIT}
\end{eqnarray}
where $\Omega_p$ ($\Omega_c$) and $\Delta_p$ ($\Delta_c$) are probe (coupling) laser Rabi frequencies and detunings, respectively. 
Without interactions with Rydberg atoms, the background gas would be transparent for the probe beam due to (Rydberg-)EIT for $\Omega_c\gg \Omega_p$, $\Delta_p=\Delta_c=0$ \cite{fleischhauer:review,friedler:longrangepulses,Mohapatra:coherent_ryd_det,mauger:strontspec,Mohapatra:giantelectroopt,schempp:cpt,sevincli:quantuminterf,parigi:interactionnonlin}. 

Van-der-Waals interactions between background and Rydberg atoms change this, and are described by
\begin{align}
	\sub{\hat{H}}{int} =&  \sum_{a\in\{s,p\},\alpha n } V_{\alpha n}^{(ua)}(\mathbf{X}_n,\mathbf{x}_\alpha) \hat{\sigma}_{uu}^{(\alpha)}\hat{\sigma}_{aa}^{(n)} ,
	\label{Hint_usp}
\end{align}
where $V_{\alpha n}^{(ua)}(\mathbf{X}_n,\mathbf{x}_\alpha)= C_{\eta(a),ua}/|\mathbf{X}_n - \mathbf{x}_\alpha |^\eta$, with $a\in\{s,p\}$ and $\eta(s)=6$, $\eta(p)=4$.  For background atoms closer than a critical distance $d_{c,a}=(C_{\eta(a),ua}\Gamma_p/\Omega_c^2)^{1/\eta(a)}$ \cite{David:Rydagg} with $a\in\{s,p\}$  to a Rydberg atom, interactions break EIT, resulting in an absorption shadow governed by $V_{\alpha n}^{(ua)}$, providing an indirect measurement of the position (and state) of the Rydberg atom through the centre of the shadow (and its radius), as shown in \fref{sketch_setup}. This is encoded in the steady state optical susceptibility
\begin{align}
 \chi_p(t) = \Gamma_p/\Omega_p\text{Im}[\hat{\rho}\sigma_{ge}^{(\alpha)}(t)],
 \label{sucept}
\end{align}
describing absorption of the probe \cite{Gunter:EIT,guenter:EITexpt,David:Rydagg}. Interactions between background atoms are negligible \cite{David:Rydagg,olmos:amplification} due to the small probability for two of them to be simultaneously excited to $\ket{u}$.

{\it Step 2:} We now add dipole-dipole interactions of the Rydberg atoms, with Hamiltonian
\begin{align}
	\sub{\hat{H}}{dd} &= \sum_{n\neq m} W_{nm}(R_{nm}) \ket{\pi_n}\bra{\pi_m}, \label{H_dd}
\end{align}
for $W_{nm}(R)=C_3/R^3$, $R_{nm}=|\mathbf{X}_n - \mathbf{X}_m|$ and strength $C_3$. The complete Hamiltonian is thus $\hat{H} = \sub{\hat{H}}{dd} + \sub{\hat{H}}{EIT} + \sub{\hat{H}}{int}$,
which enters \bref{full_main} along with $\hat{L}_\alpha$ to describe the dynamics of $\hat{\rho}$.

Now we reduce the Hilbert space dimension by adiabatically eliminating the states $\ket{e}$ and $\ket{u}$ \cite{Sorenson:eff_operator} to work with a reduced density matrix $\hat{\rho}^{(red)}=\sum_{nm}\rho_{nm}\ket{\pi_n}\bra{\pi_m}$ for the Rydberg atoms alone. If the background gas reaches its steady-state faster than dipole-dipole transitions, $W_{nm}\ll\Gamma_p$, the latter are ignored during the elimination. For slightly larger dipolar interactions $W_{nm}\approx\Gamma_p$ relevant here, we non-perturbatively include them \cite{Sorenson:eff_operator}. Either approach provides an effective  equation 
		\begin{align}
			\dot{\hat{\rho}}^{(red)}&=-i[\sub{\hat{H}}{eff}+\sub{\hat{H}}{dd},\hat{\rho}^{(red)}]+\sum_\alpha \mathcal{L}_{\sub{\hat{L}}{eff}^{(\alpha)}}[\hat{\rho}^{(red)}],\label{agg_eff} 
		\end{align}	
in the Rydberg Hilbertspace with effective operators $\sub{\hat{H}}{eff}= \sum_{m,n,\alpha} \sub{h}{eff}^{nm(\alpha)} (\mathbf{X},\mathbf{x}_\alpha)\ket{\pi_n}\bra{\pi_m}$ and $\sub{\hat{L}}{eff}^{(\alpha)}= \sum_{m,n} \sub{\ell}{eff}^{nm(\alpha)} (\mathbf{X},\mathbf{x}_\alpha)\ket{\pi_n}\bra{\pi_m}$. 

For the next step we shall write \bref{agg_eff} as $\dot{\hat{\rho}}^{(red)}={\cal A}[\hat{\rho}^{(red)}]$, defining ${\cal A}$ as the super-operator acting on the reduced density matrix, which depends on $\sub{\hat{H}}{dd}$, $\sub{\hat{H}}{eff}$ and $\sub{\hat{L}}{eff}$.

{\it Step 3:} 
We finally incorporate motion of the Rydberg atoms, at positions $\mathbf{X}$  \cite{wuester:immcrad}, extending the reduced density matrix to 
\begin{align}
	\hat{\rho}^{(red)} &= \sum_{nm}\int d\mathbf{X}d\mathbf{X'}\rho(\mathbf{X},\mathbf{X'})_{nm} \ket{\mathbf{X},\pi_n} \bra{\mathbf{X'},\pi_m}, \label{rho_xxp}
\end{align}
with $\ket{\mathbf{X},\pi_n}\equiv \ket{\mathbf{X}} \otimes \ket{\pi_n}$,
and changing $\ket{\pi_n}\bra{\pi_m}\rightarrow \ket{\mathbf{X},\pi_n}\bra{\mathbf{X},\pi_m}$ in \bref{agg_eff}.
The effective motional master equation obtained from \bref{agg_eff} is
\begin{align}
	\dot{\rho}(\mathbf{X},\mathbf{X'})_{nm} =& -\frac{i}{\hbar} \bigg[-\frac{\hbar^2}{2M}(\nabla^2_\mathbf{X}
	- \nabla^2_{\mathbf{X'}})\rho(\mathbf{X},\mathbf{X'})_{nm}\nonumber\\
	& +\sum_{kl}{\cal A}(\mathbf{X},\mathbf{X'})^{nm}_{kl}\rho(\mathbf{X},\mathbf{X'})_{kl}\bigg],\label{general_N_eom}
\end{align}
with ${\cal A}(\mathbf{X},\mathbf{X'})^{nm}_{kl}=E(\mathbf{X},\mathbf{X'})^{nm}_{kl} + \Delta E(\mathbf{X},\mathbf{X'})^{nm}_{kl} + i\gamma(\mathbf{X},\mathbf{X'})^{nm}_{kl}$. Here $ E(\mathbf{X},\mathbf{X'})^{nm}_{kl}$ involves dipole-dipole interactions $W_{nm} (\mathbf{X})$ 
and $\Delta E(\mathbf{X},\mathbf{X'})^{nm}_{kl}$ background gas effects from $\sub{h}{eff}^{nm(\alpha)}(\mathbf{X},\mathbf{x}_\alpha)$. The terms arise from the unitary part of \bref{agg_eff}, with operators projected onto coordinates $\mathbf{X}$ ($\mathbf{X'}$) if they act from the left (right) on the density matrix. Similarly, $\gamma(\mathbf{X},\mathbf{X'})^{nm}_{kl}$ accounts for the Lindblad part of \bref{agg_eff} that depends on $\sub{\ell}{eff}^{nm(\alpha)}(\mathbf{X},\mathbf{x}_\alpha)$.  We discuss the role of each element in ${\cal A}(\mathbf{X},\mathbf{X'})$ for dynamics in detail later. To reach \bref{general_N_eom}, we assume that the motion is even slower than the time-scale of dipole-dipole transitions. 

\begin{figure}[tb]
	\centering
	\epsfig{file=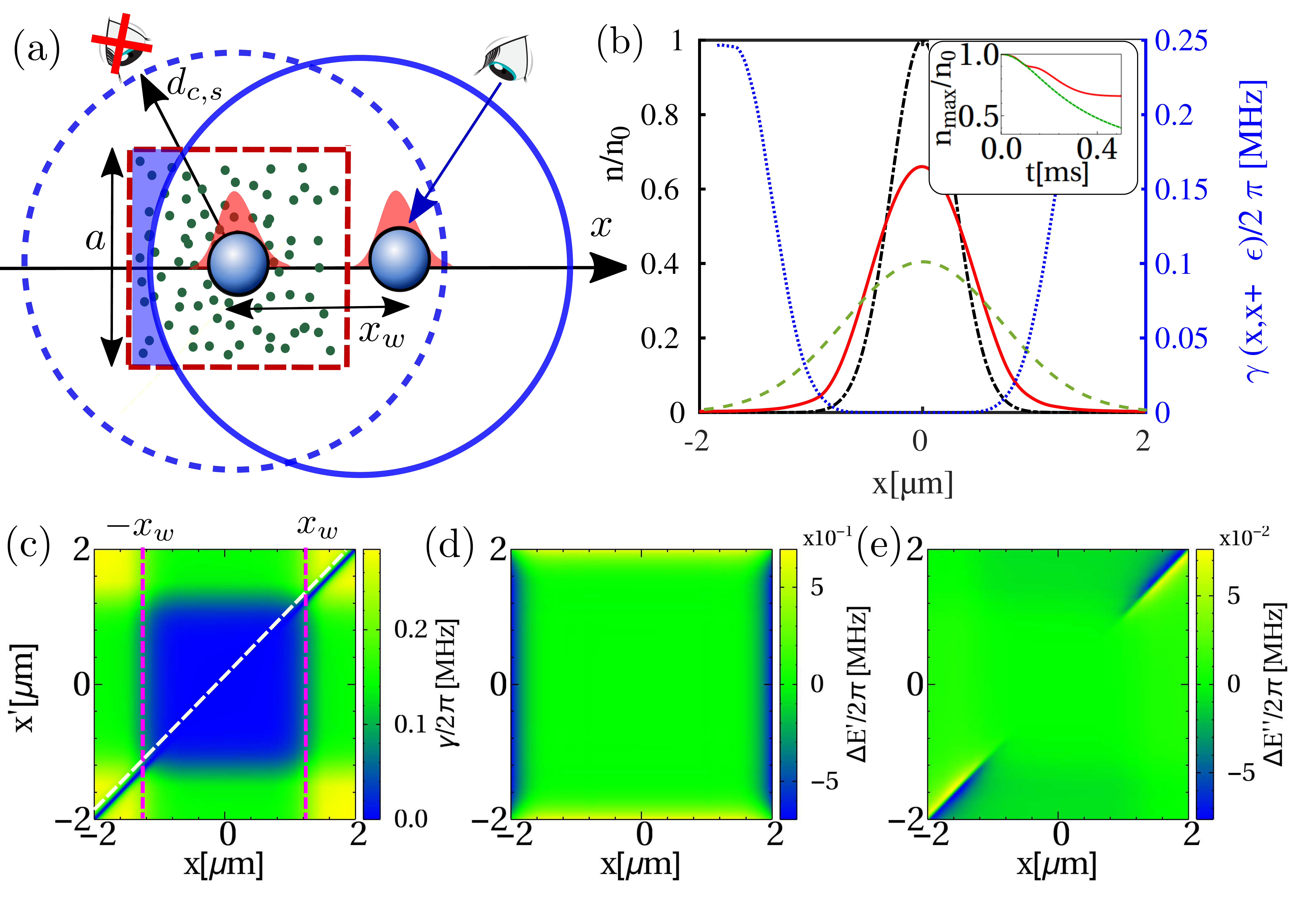,width= \linewidth}
	\caption[Caption for LOF]{Trapping a single particle through dephasing.
		(a) Sketch of atom trapped in a dephasing well resulting from position measurements  from the interrogating medium (green dots). 
		(b) Probability density $n(x)=\rho(x,x)$ of the trapped particle at the initial time $t=0$  with $\sigma=0.4\mu$m (black dot-dashed line), and $t=500$ $\mu$s later, in free space using $\gamma=0$ (green dashed) or within the dephasing well (red line). The blue dotted line shows the 1D cut $\gamma(x,x+\epsilon)$ near the diagonal, shown by the white dashed line in (b), for an offset $\epsilon=0.15$ $\mu$m. Time-evolution of the peak intensity is shown in the inset of (b).  (c)-(e) Underlying dephasing rate $\gamma(x,x')$, energy disorder $\Delta E'(x,x')$ and $\Delta E''(x,x')$ represented by \bref{gamma_xxp_main} and \bref{Delta_E_main}, respectively \cite{sup:info}. $x_w$ at the red dashed line indicates the size of the well, also sketched in (a). 
		\label{single_ryd}}
\end{figure}

\ssection{Square dephasing well}
%
We shall now demonstrate a square well due to measurement induced dephasing instead of a conservative potential. As an example, consider one Rydberg atom in the state $\ket{s}\equiv\ket{80s}$ with life-time $\tau\approx 620\mu$s at $T=0K$ \cite{beterov2009quasiclassical}. 
 EIT-based position detection using $\ket{u}\equiv\ket{31s}$ is made spatially dependent by positioning all background atoms in a cubic volume $V=a^3$ as sketched in \frefp{single_ryd}{a}. We chose the side-length $a$ significantly smaller than the radius $d_{c,s}$ of the absorption shadow near the Rydberg atom \cite{David:Rydagg}, creating a \emph{dephasing-free} region of width $2x_w$, with $x_w=d_{c,s}-a/2$, within which the optical signal gives no information on the Rydberg position, but records if the atom leaves that region.

For one Rydberg atom, $\sub{\hat{H}}{dd}=0$ in \bref{agg_eff}. 
 We set  $\mathbf{X}=[x,0,0]^T$, constraining the single Rydberg atom to one dimension for simplicity, while background atoms are distributed in 3D, with positions $\mathbf{x_\alpha}$. The effective Hamiltonian $\sub{\hat{H}}{eff}=\sub{h}{eff} (x)\ket{x}\bra{x}$ and Lindblad operator $\sub{\hat{L}}{eff}^{(\alpha)}=\sub{\ell}{eff}^{(\alpha)} (x,\mathbf{x}_\alpha)\ket{x}\bra{x}$ enters \bref{agg_eff} with \cite{David:Rydagg}
\begin{subequations}
\begin{eqnarray}
	\sub{h}{eff}(x)&=&\sum_\alpha \frac{\Omega_p^2\tilde{\Omega}_c^2\tilde{V}(x,\mathbf{x}_\alpha)}{\tilde{\Omega}_c^4+4\tilde{V}^2(x,\mathbf{x}_\alpha)(|\tilde{\Gamma}_p|^2-4\Delta_p^2)}, \label{Heff_x_main}\\
	\sub{\ell}{eff}^{(\alpha)}(x,\mathbf{x}_\alpha)&=& \frac{2i\tilde{V}(x,\mathbf{x}_\alpha)\sqrt{\Gamma_p}\Omega_p}{2\tilde{V}(x,\mathbf{x}_\alpha)\tilde{\Gamma}_p-i\Omega_c^2},
	\label{Leff_x_main}
\end{eqnarray}
\end{subequations}
where $\tilde{\Omega}_c^2=\Omega_c^2+4\tilde{V}(x,\mathbf{x}_\alpha)\Delta_p$, $\tilde{\Gamma}_p = \Gamma_p-2i\Delta_p$ and $\tilde{V}(x,\mathbf{x}_\alpha)= V^{(us)}_\alpha(\mathbf{X},\mathbf{x}_\alpha)-\Delta_p-\Delta_c$. To focus on effects from dephasing only, without a net potential, we compensate $\sub{h}{eff}(x)$ by a suitable external potential $\sub{V}{ext}(x)$, chosen such that $\sub{h}{eff}(x)+\sub{V}{ext}(x)\approx0$, see \cite{sup:info}.

Using these in \bref{general_N_eom}, we find the motional master equation for the single Rydberg atom of mass $M$ in a dephasing well as
\begin{align}
	\dot{\rho}(x,x') &= -\frac{i}{\hbar} \big[-\frac{\hbar^2}{2M}(\nabla^2_x
	- \nabla^2_{x'})\rho(x,x')\big]\nonumber\\
	& +\big(i\Delta E(x,x')/\hbar - \gamma(x,x')/2\big)\rho(x,x'),\label{eom_single}
\end{align}
with
\begin{align}
	\gamma(x,x') &= \sum_\alpha |\sub{\ell}{eff}^{(\alpha)} (x,\mathbf{x}_\alpha)|^2 + |\sub{\ell}{eff}^{(\alpha)} (x',\mathbf{x}_\alpha)|^2  \nonumber \\
	 &-2\text{Re}[\sub{\ell}{eff}^{(\alpha)} (x,\mathbf{x}_\alpha)\sub{\ell}{eff}^{(\alpha)*} (x',\mathbf{x}_\alpha)], \label{gamma_xxp_main} \\
	\Delta E(x,x') &= \Delta E'(x,x') + \Delta E''(x,x'),\label{Delta_E_main}
\end{align}
where $\Delta E'(x,x')=\sub{\tilde{h}}{eff}(x) - \sub{\tilde{h}}{eff}(x')$ and $\Delta E''(x,x')=\sum_\alpha \text{Im}[\sub{\ell}{eff}^{(\alpha)} (x,\mathbf{x}_\alpha)\sub{\ell}{eff}^{(\alpha)*} (x',\mathbf{x}_\alpha)]$, shown in \frefp{single_ryd}{c-e} with $\sub{\tilde{h}}{eff}(x)=\sub{h}{eff}(x)+\sub{V}{ext}(x)$. \eref{Leff_x_main} depends on the EIT parameters and interactions, which can both be tuned. Together with the spatial background atom distribution, this allowed creating a dephasing well shown in \fref{single_ryd}~(c). We see in the $\gamma(x,x')$ profile that dephasing will affect coherences $\rho(x,x')$ for $x$ inside and $x'$ outside the decoherence free region, or vice versa.

We initialize the Rydberg atom in a wavepacket $\psi(x)=\exp[-(x-x_0)^2/(2\sigma^2)]/(\pi \sigma^2)^{1/4}$ centered in the dephasing well, see 
black dot-dashed line in \fref{single_ryd}~(b). In free space, the wavepacket would diffuse on time-scales of interest. In the background-gas, evolving the effective Lindblad Master equation \bref{eom_single} using the high-level language XMDS \cite{xmds:paper,xmds:docu}, for parameters given in \cite{sup:info}, we show the time-evolving density in \fref{single_ryd}~(b).

We can see that spreading is strongly suppressed (red) due to the square dephasing well (blue dotted), compared to the case without dephasing $\gamma=0$ (green dashed). This could be observed by high precision Rydberg atom location measurements \cite{thaicharoen:trajectory_imaging,guenter:EITexpt}. The dephasing inhibits any further diffusion of the wavefunction beyond a critical time $t_c\approx300\mu s$, shown in inset of \fref{single_ryd}~(b). This suppression is not due to a conservative potential, since the disorder from the background $\sub{h}{eff}(x)$ has been cancelled.
Trapping arises instead solely due to the dephasing of coherences between different spatial locations $x\neq x'$. In this sense, it is decoherence due to the position measurement itself, that provides the confinement in the well. Similar ideas in another context were explored in \cite{Makrory:qmeasurement,facchi2001zeno}, for which we propose here an experimental platform using realistic parameters. The energy of the trapped atom is conserved since the environment only causes dephasing.

%
\ssection{Separation dependent dimer decoherence}
%
\begin{figure}[tb]
	\centering
	\epsfig{file=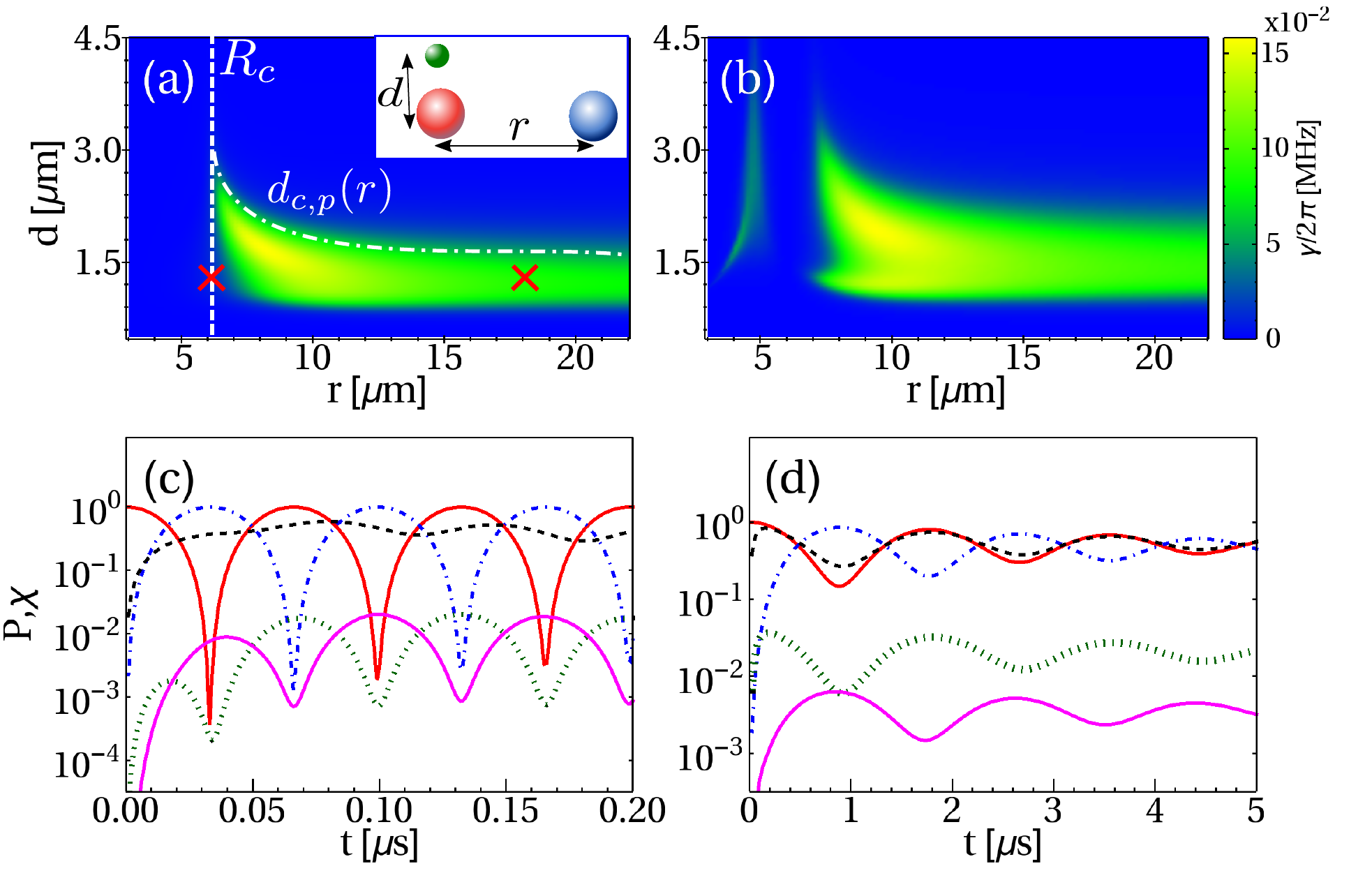,width= \figurewidth\columnwidth} 
	\caption[Caption for LOF]{Dimer decoherence ceases at close distances. (a,b) Variation of dephasing rate $\gamma(r,d)$ in a Rydberg dimer probed by a single detector atom, (a) using the three-body master equation \eref{full_main} and, (b) using the effective model \eref{agg_eff}. Parameters are in footnote \cite{para:eff_full}. The geometry is shown in the inset. (c) Population of $\ket{gps}$ (red solid), $\ket{gsp}$ (blue dot-dashed), $\ket{eps}$ (green dotted), $\ket{esp}$ (magenta solid) and optical susceptibility as described by \bref{sucept} $\chi=\kappa$Im$[\bra{g}\hat{\rho}\ket{e}]$  (black-dashed line) \cite{para:eff_full} for $r=6\mu$m and $d=1.05\mu$m. (d) same as (c) but for $r=18\mu$m. Both configuraions are marked with red crosses in (a). 
	}
	\label{full_eff_comparison}
\end{figure}
%
In order to extend the above to binding, we require the dephasing strength to depend on the \emph{relative} coordinate in a dimer
 instead of absolute coordinates. This feature is naturally provided because EIT-based Rydberg state measurements cease to be effective for very short dimer separations, for reasons discussed now. The dimer atoms are 
constrained to Rydberg states $\ket{s}\equiv\ket{43s}$ and $\ket{p}\equiv\ket{43p}$, with a life-time of $42\mu$s and $62.5\mu$s, respectively at $T=300K$ \cite{beterov2009quasiclassical}. The dimer can then be in states
	$\ket{\pi_1} = \ket{ps}$ and  $\ket{\pi_2} = \ket{sp}$ with a resultant life-time of $25\mu$s.  The background gas Rydberg state is $\ket{u}\equiv\ket{38s}$. 
	
	Now, we implement step 2 discussed earlier to derive the set of effective operators $\sub{\hat{H}}{eff}$ and $\sub{\hat{L}}{eff}$ discussed in \bref{agg_eff} for Rydberg dimer ($N=2$). The effective model developed in \rref{David:Rydagg} is not valid for close dimer separations, hence we extended it here. To keep the problem tractable, we take the relative coordinate $r$ of the dimer atoms as the only motional degree of freedom, fixing their positions in terms of $\bar{\mathbf{X}}_{1,2} = \mathbf{R}_0 \mp r \mathbf{e}_x/2$, where $\mathbf{R}_0$ is the centre of mass of dimer atoms and $\mathbf{e}_x$ a unit vector along the $x$-axis. \eref{agg_eff} in the relative coordinate $r$ contains $	\sub{\hat{H}}{eff} = \sum_{n,m,\alpha}\sub{h}{eff}^{nm(\alpha)}(r,\mathbf{x}_\alpha)\ket{\pi_n}\bra{\pi_m}$ and $\sub{\hat{L}}{eff} = \sum_{n,m,\alpha}\sub{\ell}{eff}^{nm(\alpha)}(r,\mathbf{x}_\alpha)\ket{\pi_n}\bra{\pi_m}$, which we derive in the SI \cite{sup:info}. The closed form of $\sub{h}{eff}^{nm(\alpha)}(r,\mathbf{x}_\alpha)$ and $\sub{\ell}{eff}^{nm(\alpha)}(r,\mathbf{x}_\alpha)$ are long and technical, and are inserted into our code using the export feature of Mathematica. To validate the derivation and demonstrate the cessation of decoherence for close proximity dimer atoms, we compare results from \esref{full_main}{agg_eff}, using a test case with a Rydberg dimer flanked by a single background atom as shown in the inset of \fref{full_eff_comparison}~(a), which also defines the coordinates $r$ and $d$. 
	
For panels (a,b), the dimer is initialized in $(\ket{\pi_1}+\ket{\pi_2})/\sqrt{2}$, which is evolved in time, neglecting motion, according to \esref{full_main}{agg_eff}. For all parameters, we can fit
 $\bra{\pi_1}\hat{\rho}\ket{\pi_2}$ by the exponential $\exp{[-\gamma(r,d) t]}$, from which we show the decay rate $\gamma$ in the figure. There are two main features: (i) As discussed in \cite{David:Rydagg}, the background atom only decoheres the dimer if placed at a distance $d_{c,s}(1.3\mu m)<d<d_{c,p}(2\mu m)$ from the latter. 
(ii) This decoherence ceases for dimer separations below $R_c\approx 6$ $\mu$m, where we empirically find $R_c \propto C_3^{1/3}$. Here the background can no longer adiabatically follow dipole-dipole interactions, as demonstrated in panels (c,d). These show the essential state populations together with the optical susceptibility $\chi_p$ (black dashed line) of the background atom, defined in \bref{sucept}, when the dimer is initialised in $\ket{\pi_1}$, the population of which is shown as red solid line. The background atom should show non-zero susceptibility only for dimer state $\ket{\pi_1}$, if it can adiabatically follow state changes. However the susceptibility is synchronised with the linear population oscillations only at large separation. For small $r$, the background atom can thus no longer ``measure'' the state of the dimer, and decoherence ceases. 

This creates a well in the relative coordinate $r$, based on dephasing. A dimer with initially close separation, moving towards larger distances due to constituent repulsion, will encounter an abrupt dephasing barrier in its relative coordinate. From this key result, we will show that the barrier can be strong enough to cause binding of the dimer. From panel (b), it can be seen that the effective model qualitatively captures all the features correctly, if we manually remove the decoherence feature at $r\approx 5$ $\mu$m, done in the following to tackle simulations in a background gas with $\sub{N}{bg}\approx2000$.

\ssection{Binding by dephasing}
\label{Reflection}
%
We now consider motion of the dimer atoms, as in step 3, in the regime of strong dipolar interactions, using the improved effective model benchmarked in \fref{full_eff_comparison}. The relative coordinate $r$ of the dimer atoms is the only motional degrees of freedom, restricted along the $x$-axis. Hence, the dynamic position $\mathbf{X}$ in $\hat{\rho}^{(red)}$ in \bref{rho_xxp} is now replaced by the relative coordinate $r$ of the Rydberg dimer. When deriving the master equation as in \bref{general_N_eom}, $\sub{h}{eff}^{nm(\alpha)}(r,\mathbf{x}_{\alpha})$ and $\sub{l}{eff}^{nm(\alpha)}(r,\mathbf{x}_{\alpha})$ are projected onto coordinates $r(r')$ if they act from the left (right) on the density matrix. We find
\begin{align}
	\dot{\rho}&(r,r')_{nm} = -\frac{i}{\hbar} \bigg[-\frac{\hbar^2}{2\mu}(\nabla^2_{r}
	- \nabla^2_{r'})\rho(r,r')_{nm}\nonumber\\
	& + \sum_k (W_{nk}(r)\rho(r,r')_{km} - W_{km}(r')\rho(r,r')_{nk})\bigg]\nonumber\\
	& +\sum_{k,l} \bigg(i\frac{\Delta E(r,r')^{nm}_{kl}}{\hbar} - \gamma(r,r')^{nm}_{kl}\bigg)\rho(r,r')_{kl},
	\label{general_eom}
\end{align}
where $\mu=M/2$ is the reduced mass of the dimer atoms, $W_{12}=W_{21}$ are dipole-dipole interactions \bref{H_dd} and $W_{nn}=0$. The background gas 
enters through an energy detuning $\Delta E(r,r')^{nm}_{kl}$ and dephasing $\gamma(r,r')^{nm}_{kl}$ given as
	\begin{subequations}
	\begin{eqnarray}
		\Delta E(r,r')^{nm}_{kl} &=& \sum_\alpha(\sub{h}{eff}^{nk(\alpha)}(r,\mathbf{x}_\alpha)\delta_{l,m} - \sub{h}{eff}^{lm(\alpha)}(r',\mathbf{x}_\alpha)\delta_{n,k} ) \nonumber\\
		&+&\text{Im}\big[\mathcal{O}(r,r')^{nm}_{kl}\big], \label{Delta_E_nmkl}\\
		\gamma(r,r')^{nm}_{kl} &=&\text{Re}\big[\mathcal{O}(r,r')^{nm}_{kl}\big],\label{gamma_nmkl}
		\\
		\mathcal{O}(r,r')^{nm}_{kl}&=&\sum_\alpha \bigg[[\sub{\ell}{eff}^{nk(\alpha)}(r,\mathbf{x}_\alpha)]^*\sub{\ell}{eff}^{lm(\alpha)}(r',\mathbf{x}_\alpha)\nonumber\\
		&-&\frac{1}{2}\sum_j \bigg( [\sub{\ell}{eff}^{nj(\alpha)}(r,\mathbf{x}_\alpha)]^*\sub{\ell}{eff}^{jk(\alpha)}(r,\mathbf{x}_\alpha)\delta_{l,m} \nonumber \\
		&+&[\sub{\ell}{eff}^{lj(\alpha)}(r',\mathbf{x}_\alpha)]^*\sub{\ell}{eff}^{jm(\alpha)}(r',\mathbf{x}_\alpha)\delta_{n,k}\bigg) \bigg].
\end{eqnarray}
	\end{subequations}

The first two lines of \bref{general_eom} describe unitary quantum dynamics from dipole-dipole interactions. We can diagonalize the corresponding Hamiltonian \bref{H_dd} at separation $r$ as in $\sub{\hat{H}}{dd}(r)\ket{\varphi_k}=U_k(r)\ket{\varphi_k}$ to find two Born-Oppenheimer surfaces, one repulsive and one attractive: $\sub{U(r)}{rep/att}=\pm C_3/r^3$ \cite{wuester:review}. The underlying eigenstates (molecular states) are $\ket{\sub{\varphi}{rep/att}}=(\ket{\pi_1} \pm \ket{\pi_2} )/\sqrt{2}$.

\begin{figure}[htb]
	\centering
\includegraphics[width=\columnwidth]{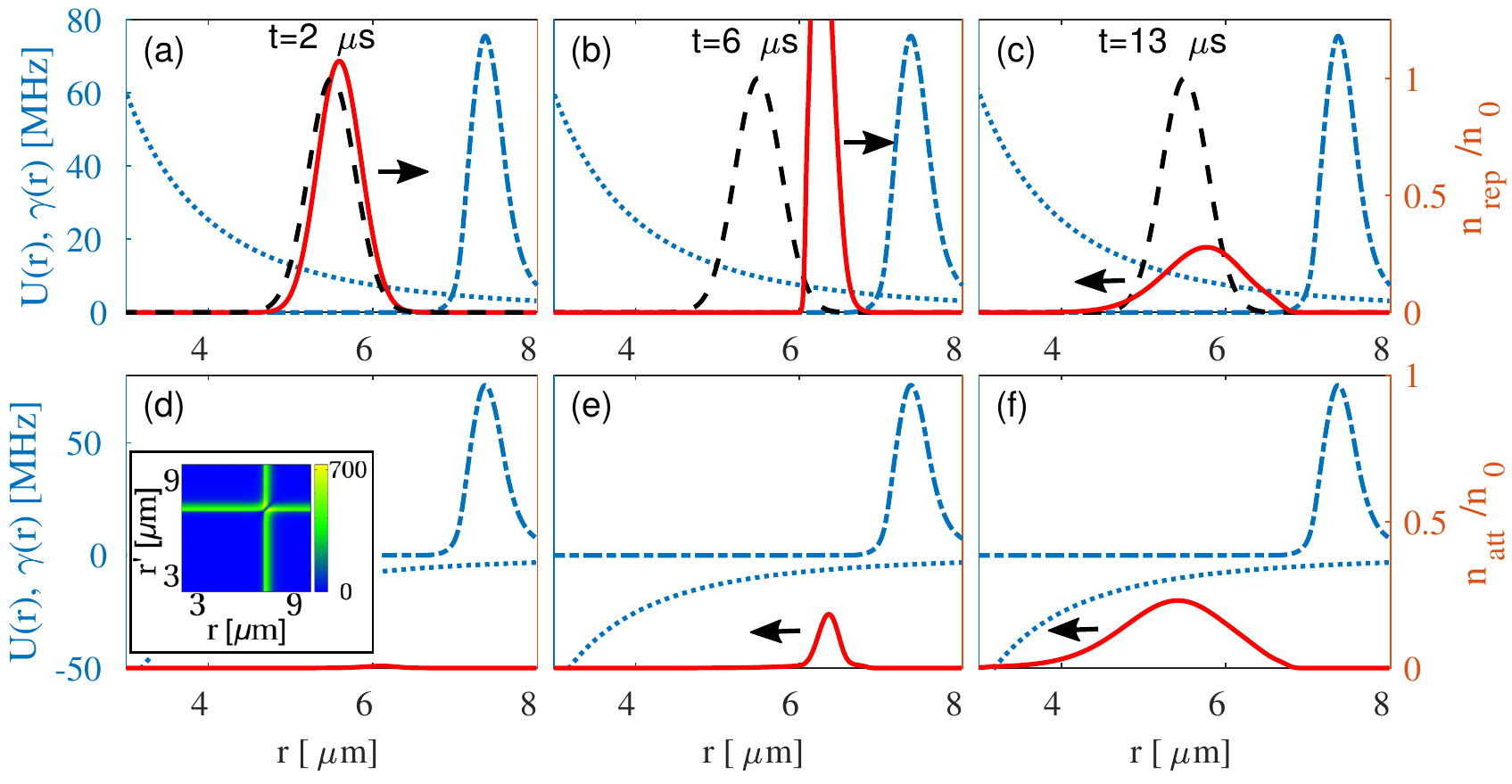}
	\caption{Dimer bound in a \emph{repulsive} potential due to reflection of a dephasing barrier. The dimer atoms are initially separated by $r = 5.5$ $\mu$m with $\sigma = 0.4$ $\mu$m and immersed in a background gas of density $\sub{\rho}{bg}=1.6\times10^{21}$ m$^{-3}$. (a)-(c) [(d)-(f)] Normalised probability density $\sub{n}{rep}/n_0$ [$\sub{n}{att}/n_0$] (solid red line) on the repulsive [attractive] surfaces (dotted blue line) at times $t=2$ $\mu$s, $6$ $\mu$s, $13$ $\mu$s. Here, $n_0$ is the initial peak density. We also show the initial densities at $t=0$ (black dashed line) and the dephasing rate $\gamma(r,r)_{12}^{12}$ (blue dot-dashed line) and $\gamma(r,r')_{nn}^{nn}$, defined in \bref{gamma_nmkl}, in the inset of (d). Arrows indicate the direction of motion of wavepackets. See supplemental material for a movie.
	\label{reflection}}
\end{figure}

 $\Delta E(r,r')^{nm}_{kl}$ represents energy shifts due to the background atoms. Here this disorder is negligible compared to the dipolar interactions. The terms $~\gamma(r,r')^{nm}_{kl}$ contain all central decoherence features. For us, two terms dominate, spatial decoherence between distances $r$ and $r'$ through $\gamma(r,r')^{nn}_{nn}$, see inset in \frefp{reflection}{d} and decoherence between electronic states $\ket{\pi_1}$ and $\ket{\pi_2}$ $\gamma(r,r')^{12}_{12}$, nearly indistinguishable on the scale of the inset except for a diagonal contribution 
$\gamma(r,r)^{12}_{12}$ shown as blue dot-dashed line in the main panels. 
At small separations ($r<R_c\approx7.5$ $\mu$m here), decoherence ceases 
since the background gas can no longer infer the dimer state. At $r$ just above $R_c$, decoherence is maximal, since for larger $r$
the critical radius $d_{c,p}(r)$ becomes smaller, reducing the volume of background atoms that contribute.

We initialise the dimer in the repulsive electronic state with a Gaussian relative wavefunction $\hat{\rho}(t=0) =\ket{\phi_0}\ket{\sub{\varphi}{rep}}\bra{\sub{\varphi}{rep}}\bra{\phi_0}$,
with $\braket{r}{\phi_0}=\phi_0(r)$=$\exp{[-(r-r_0)^2/(2\sigma^2)]}/(\pi\sigma^2)^{1/4}$.

From this initial state, we numerically solve \bref{general_eom}. 
As shown in \fref{reflection}, the Rydberg dimer on the repulsive surface reaches the dephasing barrier $\gamma(r)$ at larger dimer separation $r$ and reflects from it with a probability $\rho_{rep}\approx 47\%$ to subsequently climb the repulsive potential $U(r)$ again. The reflected part of the dimer wavepacket continues to oscillate on the repulsive energy surface, forming a decoherence induced metastable bound state. Without binding by dephasing, the wavepacket would reach $r>10 \mu$m by time $t=13 \mu$s and the dimer would dissociate. Besides the reflected component on the repulsive surface, we see in \frefp{reflection}{f} a comparable fraction incoherently coupled to the attractive surface \cite{wuester:immcrad}. $\rho_{rep}$ can be experimentally controlled by varying the strength of the dephasing barrier through EIT parameters. The binding on the repulsive surface could be seen in an experiment at e.g. $t=20\mu$s, when the population on the attractive surface would have collided and ionized, while the reflected wavepacket on the repulsive surface reached its inner turning point instead of dissociation. The separation dependence of $\gamma(r)$ is set through a competition between dipole-dipole interaction and medium response time-scales, and does not require any spatially selective optics. Similar counter-intuitive bound-states despite repulsive interactions have been reported for e.g.~repulsive Fermions in optical lattices, which can be bound due to the lattice band-gap \cite{Fallani:united_rep,Winkler:rep_bound,Petrosyan:qliquid}. While there is some minor dissipation in the scenario of \fref{reflection}, it plays no role in the binding process, as we have verified with a simulation without the terms in \bref{general_eom} that cause it. We confirmed that the dynamics in \fref{reflection} is nearly unchanged if $\Delta E(r,r')^{nm}_{kl}$ and components of $\gamma(r,r')^{nm}_{kl}$ not discussed above are neglected, but show figures of all of these in the SI \cite{sup:info}. 

\ssection{Conclusions and outlook}
%
We have shown how two central features of nature, trapped states and bound states, can arise solely based on dephasing spatial or electronic quantum coherence. Dephasing can be controlled through measurements of absolute or relative coordinates. 

This significantly extends earlier reports of binding by particle loss \cite{Lemeshko:binding,Syassen:dissipation,Syassen:dissipation_exp,Ates:diss_binding,Cui:distance_dissi,Zhu:loss_zeno,Daley:3body_loss,Vuglar_nonconsforces_PhysRevLett}, generalizing them to the most widespread type of decoherence and removing the need for dissipation. We discussed both features with examples that should be within reach of state-of-the-art 
 experiments, in which a Rydberg atom is embedded in a cold EIT medium, and the position and electronic states of the former can be controllably decohered by the latter \cite{Gunter:EIT,guenter:EITexpt,David:Rydagg,schempp:spintransport,wuester:immcrad}.
Ultimately, binding or trapping through dephasing may give rise to new states of quantum matter, similar to 
the dissipative stabilization of a Mott insulator \cite{ma2019dissipatively}, but removing the need to compensate loss.

\begin{acknowledgments}
We gladly acknowledge interesting discussions with Sidharth Rammohan, Aparna Sreedharan, Shivakant Tiwari, Shannon Whitlock and thank the Max-Planck society for financial support under the MPG-IISER partner group program as well as the Indo-French Centre for the Promotion of Advanced Research - CEFIPRA. K.M.~acknowledges the Ministry of Education for the Prime Minister's Research Fellowship (PMRF).
\end{acknowledgments}
\appendix
\beginsupplement
\begin{center}
	
	\textbf{Supplemental material}
	
\end{center}

\section{Effective model for EIT imaging with corrections due to dipole-dipole interaction}
%
The starting point for the treatment of the Rydberg impurities in the cold gas is already discussed in detail in the main text. Here, we will derive the effective Hamiltonian $\sub{\hat{H}}{eff}$ and effective Lindblad operator $\sub{\hat{L}}{eff}$ discussed in \eref{agg_eff}. It was described in \cite{David:Rydagg}, how one can use the effective operator method of \cite{Sorenson:eff_operator} to obtain a master equation for the electronic state of the dimer only, through adiabatic elimination of the dynamics in the sector $\{e,r\}$ of the background.

A physical effect that can be neglected for parameters chosen in those works, but not here, is that dipole-dipole interactions can counter decoherence. To leading order, we can incorporate this effect through the extended formalism discussed in \cite{Sorenson:eff_operator}, in which dipole-dipole interactions (more generally dynamics in the system space of interest), are taken into account when performing the adiabatic elimination of the background. The resultant reduction of decoherence rates at short dimer distances, where dipole-dipole interactions are strong, is central to the present results. In this section, we will thus describe the steps for obtaining the effective Hamiltonian $\hat{H}_\text{eff}$ and the effective decay  operators $\hat{L}^{(\alpha)}_\text{eff}$ for a Rydberg assembly consisting of $N$ atoms (for the present work $N=2$), while non-perturbatively including dipole-dipole interactions.

The complete Hamiltonian of the system is given in the form
\begin{equation}
	\hat{H} = \sub{\hat{H}}{dd} + \sub{\hat{H}}{EIT} + \sub{\hat{H}}{int},
\end{equation}
where $\sub{\hat{H}}{EIT}$, $\sub{\hat{H}}{int}$ and $\sub{\hat{H}}{dd}$ are described in \bref{H_EIT}, \bref{Hint_usp} and \bref{H_dd} of the main text.

While our division of the complete Hamiltonian above was according to the physics described, for the adiabatic elimination we divide it slightly differently into four parts according to
\begin{equation}
	\hat{H}=\hat{H}_g + \hat{H}_e + \hat{V}_+ + \hat{V}_-,
	\label{Htot2}
\end{equation}
where $\hat{H}_g$ ($\hat{H}_e$) is the Hamiltonian in the ground (excited) subspace and $\hat{V}_- (\hat{V}_+)$ are the parts of the Hamiltonian coupling these.
The splitting shall be such that dynamics in $\hat{H}_g$ is of interest, while that in $\hat{H}_e$ is to be adiabatically eliminated. In \bref{Htot2}, $\hat{H}_a$ is defined as $\hat{H}_a \equiv \hat{P}_a\hat{H}\hat{P}_a$ for $a\in\{g,e\}$, where $\hat{P}_g$ and $\hat{P}_e$ are the projection operators onto the ground and excited state, respectively, expressed as $\hat{P}_g = \sum^{N}\limits_{n=1} \ket{\boldsymbol{g}}\bra{\boldsymbol{g}}\otimes \ket{\pi_n}\bra{\pi_n}$ and $\hat{P}_e = 1 - \hat{P}_g$ with $\ket{\boldsymbol{g}}=\ket{gg...g...g}$ denoting a many-body state where all background atoms are in the $\ket{g}$ state. In the adiabatic elimination of the $\ket{e}$-space discussed in \rref{Sorenson:eff_operator}, the coupling between two subspaces $\hat{V}_+ \equiv P_e\hat{H}P_g$ and $\hat{V}_- \equiv P_g\hat{H}P_e$ is assumed to be perturbative. Here it takes the form
\begin{eqnarray}
	\hat{V}_+ = \frac{\Omega_p}{2}\sum_{n=1}^{N}\sum_\alpha[\ket{e}\bra{g}]_\alpha \otimes\ket{\pi_n}\bra{\pi_n},\\
	\hat{V}_- = \frac{\Omega_p}{2}\sum_{n=1}^{N}\sum_\alpha[\ket{g}\bra{e}]_\alpha\otimes\ket{\pi_n}\bra{\pi_n}.
\end{eqnarray}
Ref.~\cite{Sorenson:eff_operator} then provides a recipe to non-perturbatively include the effect of $\hat{H}_g$, i.e.~$W_{nk}$. Following that, we first diagonalize the groundstate Hamiltonian $\hat{H}_g$ as
\begin{equation}
	\hat{H}_g=\sum\limits_{l=1}^{N}E_l\ket{\varphi_l}\bra{\varphi_l} \equiv \sum\limits_{l=1}^{N} E_l P_l,
	\label{H_g}
\end{equation}
where $E_l$ is the eigenstate energy and $P_l$ is the corresponding projection operator $P_l = \ket{\varphi_l}\bra{\varphi_l}$. The coupling operators $\hat{V}_\pm$ are then also decomposed in terms of the eigenstates, 
\begin{align}
	\hat{V}_+ &= \sum_l \hat{V}^l_+ , \mbox{    with} 	\label{V+l}\\
	\hat{V}^l_+&=  \bigg[\frac{\Omega_p}{2}\sum_{n=1}^{N}\sum_\alpha[\ket{e}\bra{g}]_\alpha\otimes\ket{\pi_n}\bra{\pi_n}\bigg]\ket{\varphi_l}\bra{\varphi_l},\nonumber
\end{align}

where we have defined $\hat{V}^l_+$ as $\hat{V}^l_+ \equiv \hat{V}_+ P_l$, describing the excitation from the eigenstate $\ket{\varphi_l}$.

The effective Hamiltonian $\hat{H}_\text{eff}$ and decay operators $\hat{L}^{(\alpha)}_\text{eff}$ described in \cite{Sorenson:eff_operator} then are
\begin{eqnarray}
	\sub{\hat{H}}{eff} &=& -\frac{1}{2}\bigg[\hat{V}_- \sum_l \big(\sub{\hat{H}^{(l)}}{NH}\big)^{-1} \hat{V}^l_+ + \text{H.c.}\bigg] + \hat{H}_g,\label{Heff}\\
	&=& \sum_{nm,\alpha} \sub{H}{eff}^{nm(\alpha)}(\mathbf{X},\mathbf{x}_\alpha) \ket{\pi_n}\bra{\pi_m},\\
	\sub{\hat{L}}{eff}^{(\alpha)} &=& \hat{L}_\alpha \sum_l \big(\sub{\hat{H}^{(l)}}{NH}\big)^{-1} \hat{V}^l_+,
	\label{Leff}\\
	&=& \sum_{nm} \sub{L}{eff}^{nm(\alpha)}(\mathbf{X},\mathbf{x}_\alpha) \ket{\pi_n}\bra{\pi_m},
\end{eqnarray}
where $\big(\sub{\hat{H}^{(l)}}{NH}\big)^{-1} \equiv \sub{(\hat{H}}{NH} - E_l)^{-1}$ is the inverse of the non-Hermitian Hamiltonian for each eigenstate $\ket{\varphi_l}$. In that expression $\sub{\hat{H}}{NH}$ is the non-Hermitian Hamiltonian
\begin{equation}
	\sub{\hat{H}}{NH} = \hat{H}_e -\frac{i}{2}\sum_\alpha \hat{L}_\alpha^\dagger\hat{L}_\alpha,
	\label{Hnh}
\end{equation}
where $\hat{H}_g$ and $\hat{H}_e$ are defined after \bref{Htot2}. For 
$\Delta_{p,c}=0$, they can be written as
\begin{eqnarray}
	\hat{H}_g &=& \sum_{n\neq m} W_{nm}(R_{nm})\ket{\boldsymbol{g}}\bra{\boldsymbol{g}}\otimes\ket{\pi_n}\bra{\pi_m}, \\
	\hat{H}_e&=& \sum_{n,\alpha} \bigg\{ \frac{\Omega_c}{2}\bigg([\ket{e}\bra{u}]_\alpha+[\ket{u}\bra{e}]_\alpha\bigg)\ket{\pi_n}\bra{\pi_n} \nonumber\\
	&+& \sum_{\substack{a\in \{e,r\}\\m\neq n}}\bigg[ W_{nm}(R_{nm}) \big[\ket{a}\bra{a}\big]_\alpha\ket{\pi_n}\bra{\pi_m}\bigg] \nonumber\\
	&+& \tilde{V}_{\alpha n}(\mathbf{X},\mathbf{x}_\alpha)[\ket{u}\bra{u}]_\alpha \ket{\pi_n}\bra{\pi_n} \bigg\} ,
	\label{H_e}
\end{eqnarray}
where $\tilde{V}_{\alpha n}(\mathbf{X},\mathbf{x}_\alpha) = V^{(rp)}_{\alpha n}(\mathbf{X}_n,\mathbf{x}_\alpha) + \sum\limits_{m\neq n} V^{(rs)}_{\alpha m}(\mathbf{X}_m,\mathbf{x}_\alpha)$. Inserting Eqns.~\bref{H_g}, \bref{V+l} and \bref{H_e} into \bref{Heff} and \bref{Leff} leads us to the final expressions for $\sub{H}{eff}^{nm(\alpha)}(\mathbf{X},\mathbf{x}_\alpha)$ and $\sub{L}{eff}^{nm(\alpha)}(\mathbf{X},\mathbf{x}_\alpha)$, respectively. 
The dependence of the effective operators on all coordinates $\mathbf{X}$ and $\mathbf{x}$, renders the treatment of any motional dynamics in a dense background gas computationally prohibitive. To simplify this scenario, we have employed two approximations: (i) we neglect motion of the background atoms, allocating them static positions, and (ii) we take the relative coordinate $r$ of the dimer atoms as the only motional degree of freedom.

\section{External potential to counter energy disorder}
\begin{figure}[htb]
	\centering
	\epsfig{file=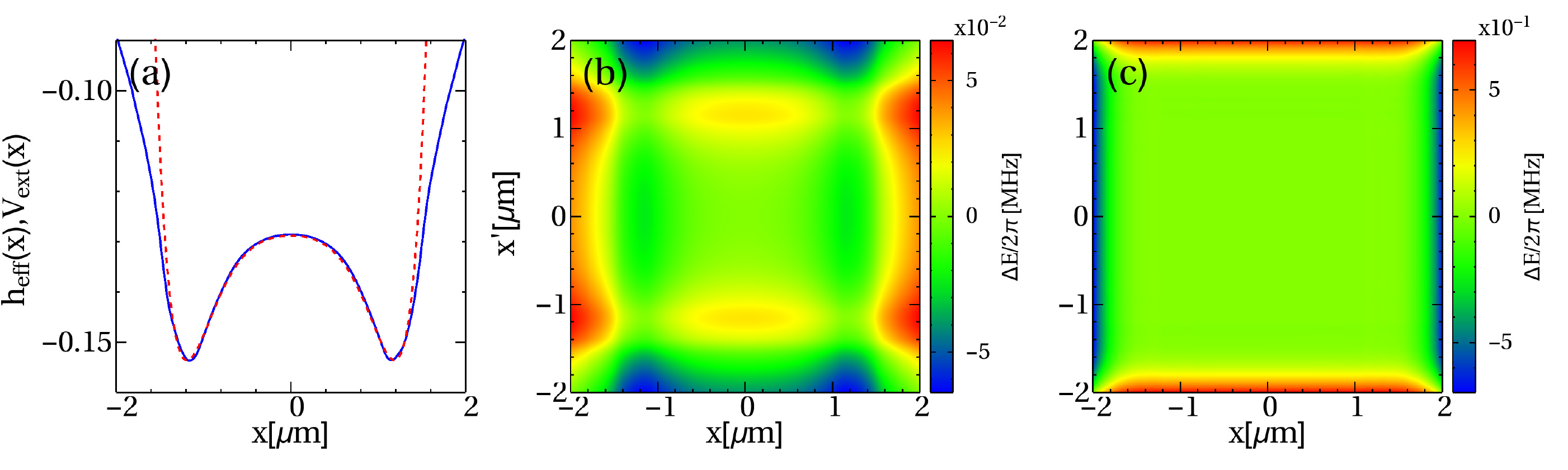,width= 0.99\columnwidth} 
	\caption{ (a) Comparison of $-\sub{h}{eff}(x)$ (blue solid), defined \eref{Heff_x_main} of the main text, and external potential $\sub{V}{ext}(x)$ (red dashed) used to compensate effects of $\sub{h}{eff}$ given in \bref{Vext}. Distribution of disorder $\Delta E'(x,x')$ defined in \eref{Delta_E_main}, (b) using $\sub{h}{eff}(x)$ and (c) using $\sub{\tilde{h}}{eff}(x)=\sub{h}{eff}(x)+\sub{V}{ext}(x)$ for EIT parameters $\Omega_p/2\pi=0.05$ MHz, $\Omega_c/2\pi = 5$ MHz, $\Gamma_p/2\pi = 6.1$ MHz, $\Delta_p=-\Delta_c=50$ MHz and $C_6/2\pi  = -88$ MHz $\mu$m$^6$ for van-der-Waals interactions between  $\ket{s}\equiv\ket{80s}$ and $\ket{u}\equiv\ket{31s}$ \cite{arc}. There are $\sub{N}{bg}=10^4$ background atoms in a 3D cubic box of sidelength $a=1.5$ $\mu$m, corresponding to a density of $\sub{\rho}{bg}=2.96\times10^{21}$ m$^{-3}$. }
	\label{Dephase_single}
\end{figure}

\begin{figure}[tb]
	\centering
	\epsfig{file=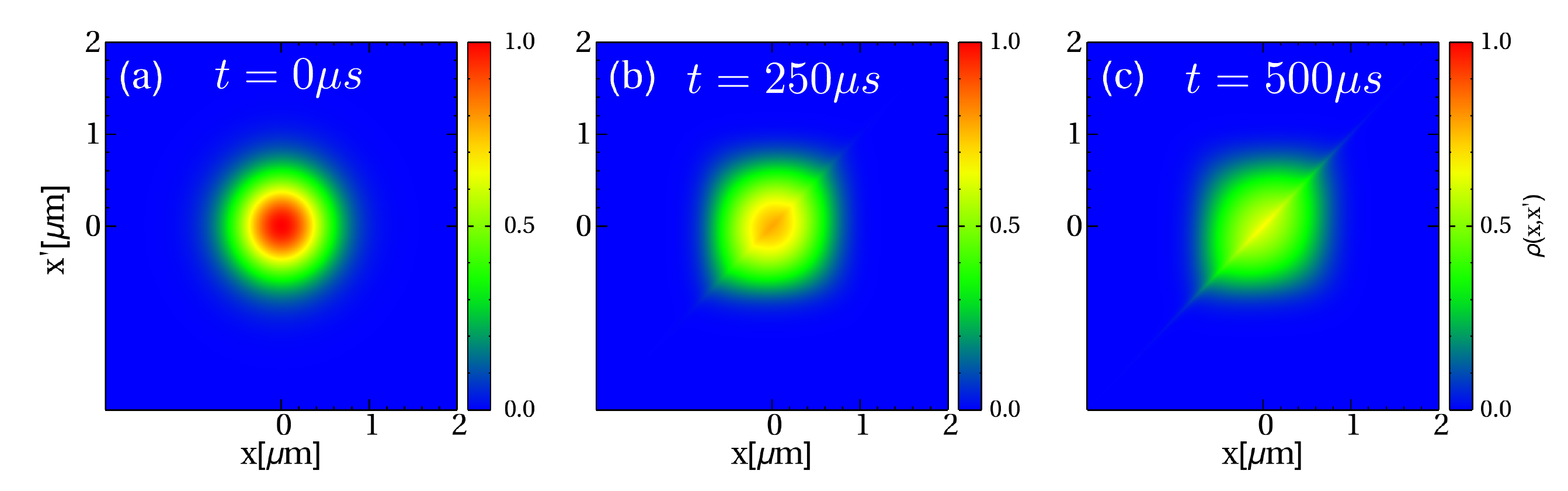,width= 0.99\columnwidth} 
	\caption{ Spatial coherence $\rho(x,x')$ corresponding to Fig.~2(b) (red-line) of the main text for particle trapped in a dephasing well.}
	\label{2D_single}
\end{figure}

%
In the setting of single Rydberg atom trapped through the dephasing induced by background atom interactions, discussed in the main text, we assume an additional external double-well potential for the Rydberg atom to compensate the Hamiltonian $\sub{\hat{H}}{eff}$, defined in  \eref{Heff_x_main} of the main text. This ensures the contributions from $\sub{\hat{H}}{eff}$ does not result in the trapping of the atom. The external potential is given in the form
\begin{eqnarray}
	\sub{V}{ext}(x)&=&a+bx^2+cx^4+dx^6+ex^8,
	\label{Vext}
\end{eqnarray}
where the coefficients are given by ($a=-0.12885$ MHz, $b=-0.0095$ MHz$/\mu$m$^2$, $c=-0.0164$ MHz$/\mu$m$^4$, $d=0.0023$ MHz$/\mu$m$^6$ and $e=0.00374$ MHz$/\mu$m$^8$) were obtained through curve-fitting into $-\sub{h}{eff}(x)$. The external double-well potential trap is shown in \fref{Dephase_single}~(a) along with it's comparison with $-\sub{h}{eff}(x)$. In \fref{Dephase_single}~(b), we show the distribution of disorder $\Delta E'(x,x')$ defined in \bref{Delta_E_main} of the main text, with $\sub{h}{eff}(x)$ defined in \bref{Leff_x_main}, while in \fref{Dephase_single}~(c), we show the distribution in (b) after incorporating the external potential. We also show the evolution of the density of the system in \fref{2D_single}.

%

\section{Spatial distribution of disorder and dephasing in dimer}
\begin{figure}[htb]
	\centering
	\epsfig{file=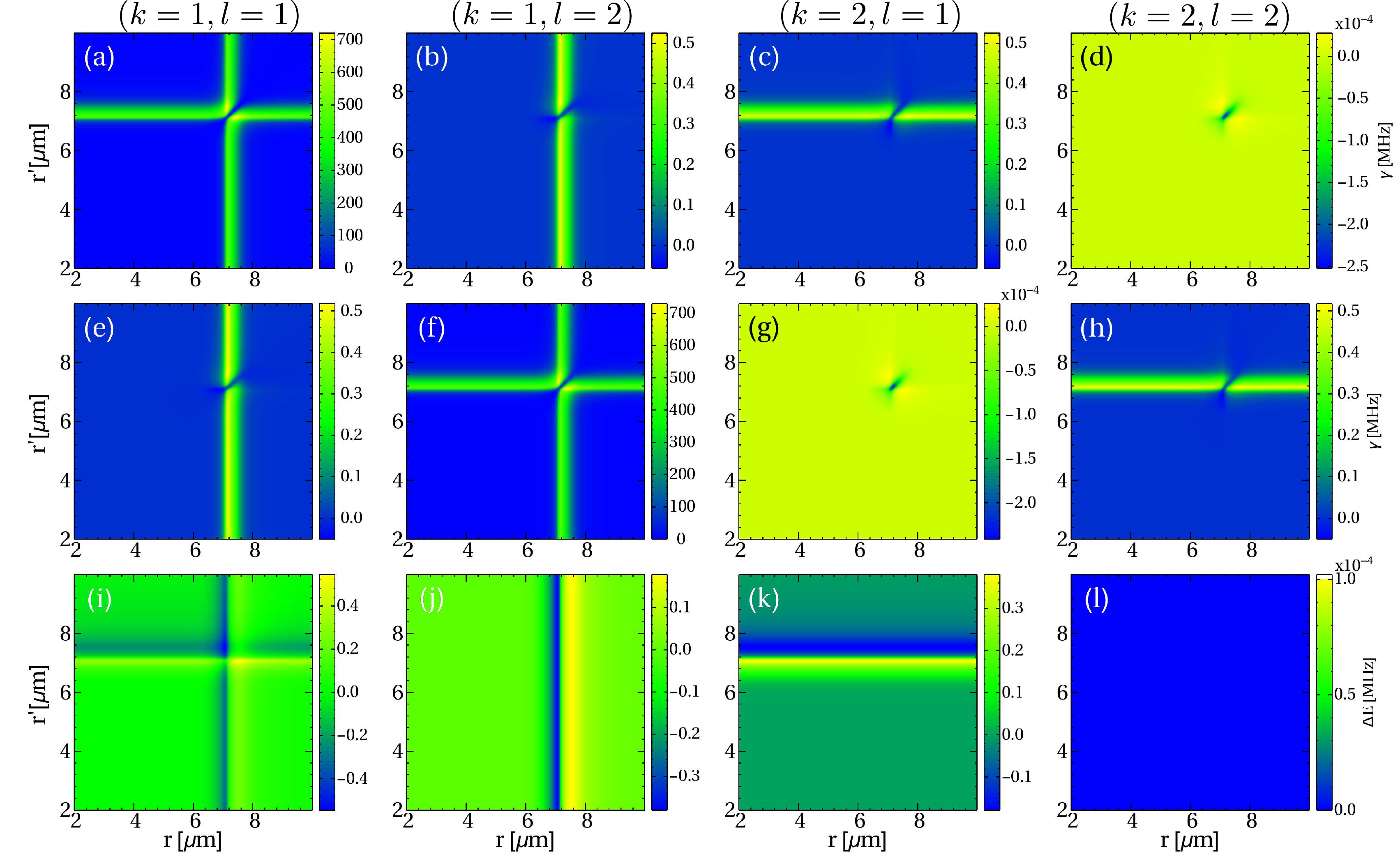,width=0.99 \linewidth}
	\caption{(a)-(h) Distribution of $\gamma(r,r')_{kl}^{nm}$ from \bref{gamma_nmkl} for $n=1$ and $m=1$ in (a)-(d), $n=1$ and $m=2$ in (e)-(h) for parameters $\Omega_p=2.3$ MHz, $\Omega_c=30$ MHz, $\Gamma_p=1.35$ MHz, $\Delta_p=\Delta_c=0$ MHz, $\sub{N}{bg}=2000$ atoms and $\rho_{bg}=1.6\times 10^{21}$ m$^{-3}$. (i)-(l) Distribution of $\Delta E(r,r')_{kl}^{nm}$ from \bref{Delta_E_nmkl} in the main text for $n=1$ and $m=1$ for the same parameters as (a)-(h). For all other $n$, $m$ values, the plots are the same or can be obtained using a conjugate transpose.
		\label{SL_coeff}}
\end{figure}

%
In this section, we will present details on the terms entering the equation of motion of the dimer,
\begin{align}
	\dot{\rho}&(r,r')_{nm} = -\frac{i}{\hbar} \bigg[-\frac{\hbar^2}{2M}(\nabla^2_{r}
	- \nabla^2_{r'})\rho(r,r')_{nm}\nonumber\\
	& + \sum_k (W_{nk}(r)\rho(r,r')_{km} - W_{km}(r')\rho(r,r')_{nk})\bigg]\nonumber\\
	& +\sum_{k,l} \bigg(i\frac{\Delta E(r,r')^{nm}_{kl}}{\hbar} - \gamma(r,r')^{nm}_{kl}\bigg)\rho(r,r')_{kl},\label{eom_SI}
\end{align}
which we have derived in the main text \bref{general_eom}. As discussed in the main text, the first two terms describe the kinetic energy of the Rydberg dimer with $r$ and $r'$ describing a single relative coordinate in the density matrix, while the third and fourth terms are dipole-dipole interactions. $\Delta E(r,r')^{nm}_{kl}$ denotes the disorder  in energy present in the system and $\gamma(r,r')^{nm}_{kl}$ is the dephasing term arising from the measurement induced decoherence. In \fref{SL_coeff}, we show the distribution of $\gamma(r,r')^{nm}_{kl}$ and $\Delta E(r,r')^{nm}_{kl}$ for $N=2$, $\sub{N}{bg}=2,000$ or $\rho_{bg}=1.6\times 10^{21}$ m$^{-3}$.  
\section{Surface densities}
%
\begin{figure}[htb]
	\centering
	\epsfig{file=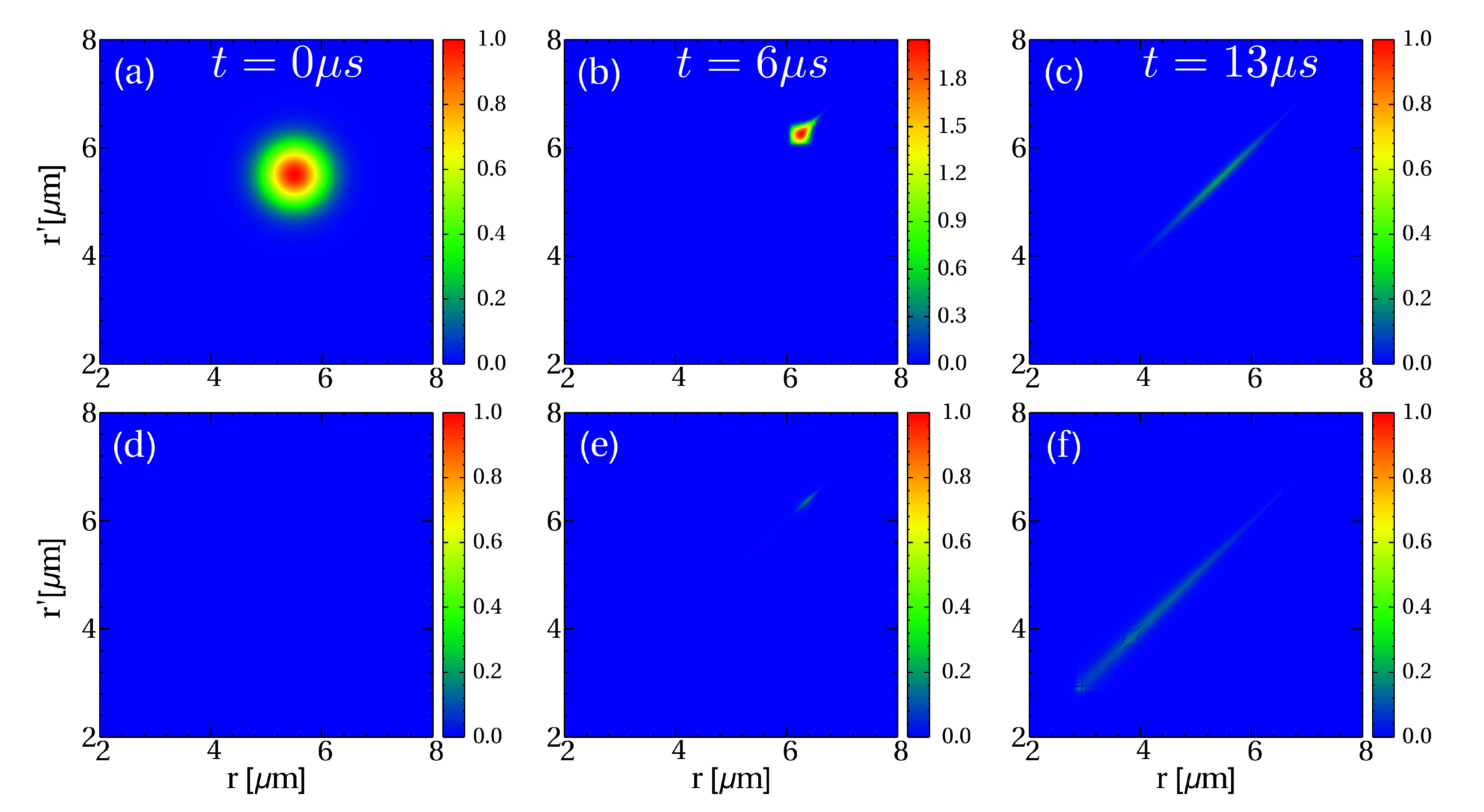,width=0.99 \linewidth}
	\caption{Surface segregated density matrix elements $\sub{\rho(r,r')}{rep}$ and $\sub{\rho(r,r')}{att}$ defined in \bref{rho_rep} and \bref{rho_att}, respectively, for the same scenario as in Fig. 4 of the main text.
		\label{density_plot}}
\end{figure}
In the main article, we numerically solve \bref{eom_SI} for a Rydberg dimer initially moving on the repulsive surface with the dephasing profile and energy disorder shown in \frefp{SL_coeff}{a-h} and \frefp{SL_coeff}{i-l}. We then plot the population of repulsive and attractive surfaces in Fig.~4 of the main article to show how the incorporation of a steep dephasing induced loss leads to the formation of a metastable bound state. For that, the population on the repulsive and attractive surfaces can be found from
\begin{eqnarray}
	\sub{n}{rep}(r) &=& \frac{1}{2}\int [\rho(r,r')_{11} + \rho(r,r')_{22} \nonumber\\
	& &+ \rho(r,r')_{12} + \rho(r,r')_{21}]dr',\\
	\sub{n}{att}(r) &=& \frac{1}{2}\int [\rho(r,r')_{11} + \rho(r,r')_{22} \nonumber \\
	& &- \rho(r,r')_{12} - \rho(r,r')_{21}]dr',
\end{eqnarray}
where $\rho(r,r')_{11}$ and $\rho(r,r')_{22}$ denotes the population of the states $\ket{\pi_1}$ and $\ket{\pi_2}$ respectively and the other terms are the coherences between these two states. The corresponding density matrix elements $\sub{\rho}{rep}(r,r')$ and $\sub{\rho}{att}(r,r')$ are written as 
\begin{eqnarray}
	\sub{\rho}{rep}(r,r') &=&\frac{1}{2}\bigg(\rho(r,r')_{11} + \rho(r,r')_{22} \nonumber\\
	& &+ \rho(r,r')_{12} + \rho(r,r')_{21}\bigg), \label{rho_rep} \\
	\sub{\rho}{att}(r,r') &=&\frac{1}{2}\bigg(\rho(r,r')_{11} + \rho(r,r')_{22} \nonumber\\
	& &- \rho(r,r')_{12} - \rho(r,r')_{21}\bigg), \label{rho_att}
\end{eqnarray}
and are shown in \fref{density_plot}.

%

\end{document}